\documentclass[a4paper,11pt]{article}

\usepackage{url}
\usepackage{latexsym}
\usepackage[small,bf]{caption}
\usepackage{amsfonts}
\usepackage{amssymb}
\usepackage{amsmath}
\usepackage{stmaryrd}
\usepackage{bbm}
\usepackage{times}
\usepackage{graphicx}
\usepackage{hyperref}

\newcommand{\Z}{{\mathbb{Z}}}

\begin{document}

%
%

\title{Pattern formation through genetic drift at expanding population fronts}

\author{Adnan Ali\footnotemark[1], Stefan Grosskinsky\footnotemark[1]}

%
%
%

\maketitle


\begin{abstract}
We investigate the nature of genetic drift acting at the leading edge of range expansions, building on recent results in [Hallatschek et al., Proc.\ Natl.\ Acad.\ Sci., \textbf{104}(50): 19926 - 19930 (2007)]. A well mixed population
of two fluorescently labeled microbial species is grown in a circular geometry. As the population expands, a coarsening process driven by genetic drift gives rise to sectoring patterns with fractal boundaries, which show a non-trivial asymptotic distribution. Using simplified lattice based Monte Carlo simulations as a generic caricature of the above experiment, we present detailed numerical results to establish a model for sector boundaries as time changed Brownian motions. This is used to derive a general one-to-one mapping of sector statistics between circular and linear geometries, which leads to a full understanding of the sectoring patterns in terms of annihilating diffusions.\\
\\
\textbf{keywords}: genetic drift; range expansion; domain coarsening; time-changed Brownian motion; annihilating random walks
\end{abstract}

\section{Introduction}

\footnotetext[1]{Centre for Complexity Science, University of Warwick, Coventry CV4 7AL, UK}
Genetic drift is one of several evolutionary processes which lead to changes in allele frequencies in populations over time. As opposed to natural selection, it is driven purely by unspecific random sampling and chance events, and is now widely accepted to be a major evolutionary force (see \cite{3h,1} and references therein). In addition to selective pressure, genetic drift leads to variations in allele frequencies. In large populations the variations are typically small compared to the population size, and these systems are dominated by the law of large numbers. But in small populations genetic variation and fluctuations in allele frequencies can have a significant impact \cite{2h} and may even lead to speciation \cite{3h}. For some species such as humans it has even been shown to be the primary force responsible for a particular feature \cite{3,4}. Range expansions are frequently seen in the history of many species \cite{6} including humans \cite{7}, when they encompass new areas to draw upon greater resources in the newly found habitats or driven by changing environmental conditions. Climatic changes trigger episodes of range contractions and expansions in many plant and animal species. For example, in recent decades, the increase in global temperatures has lead to pole-ward range extensions for a variety of species \cite{9}. Range expansions leave a distinct pattern in the genetic structure of a population, which is found to be a reduction in diversity and an increased linkage between genetic markers \cite{6}. This enables historical patterns of range expansions to be traced \cite{7}. A special case of genetic drift and range expansion is known as the founder effect. When a small subset of the population breaks away and forms a new colony, the founders have strong effects on the newly created population's genetic make-up. Through random sampling it is expected that there will be differences in allele frequencies between the original population and the colony, and over the course of many generations this genetic difference will amplify \cite{5}. For example the expansion of humans from a small population in Africa to the whole world has lead to a decrease in heterozygosity in the population \cite{3,6,7,8}.

\begin{figure}
\begin{center}
\raisebox{5mm}[0mm]{\includegraphics[height=1.7in]{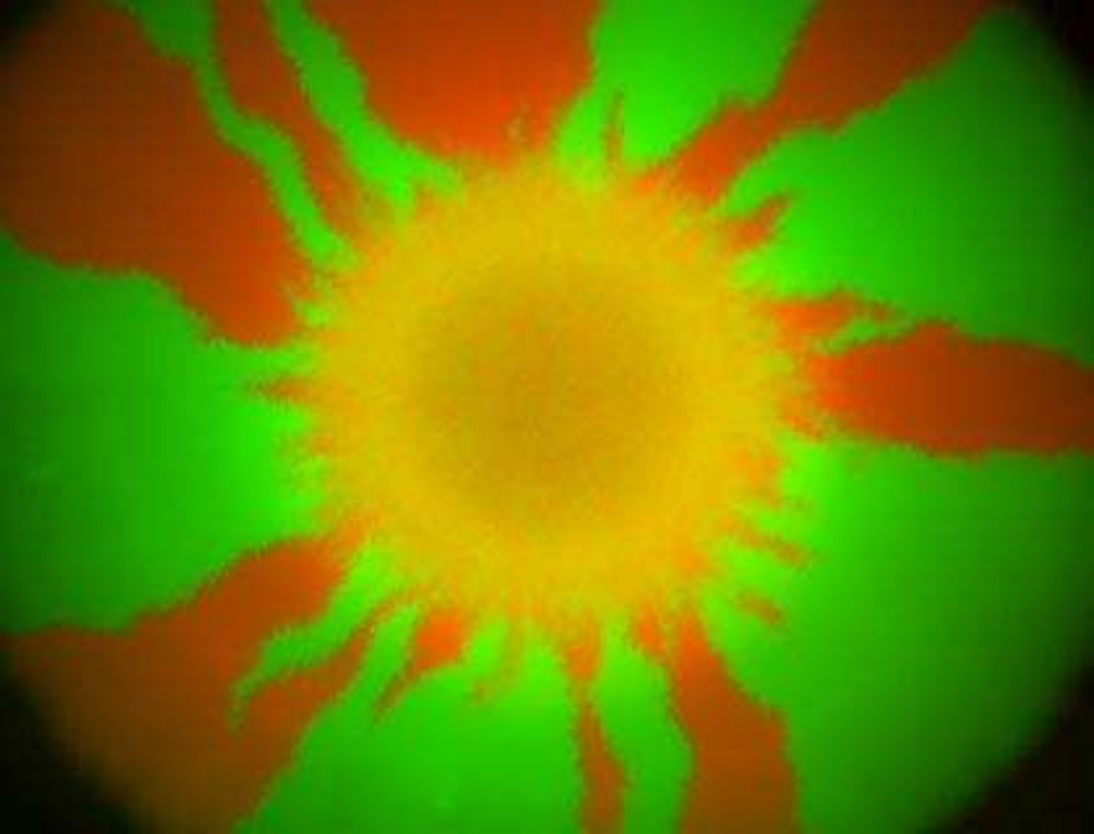}}\qquad \includegraphics[height=2in,width=2in]{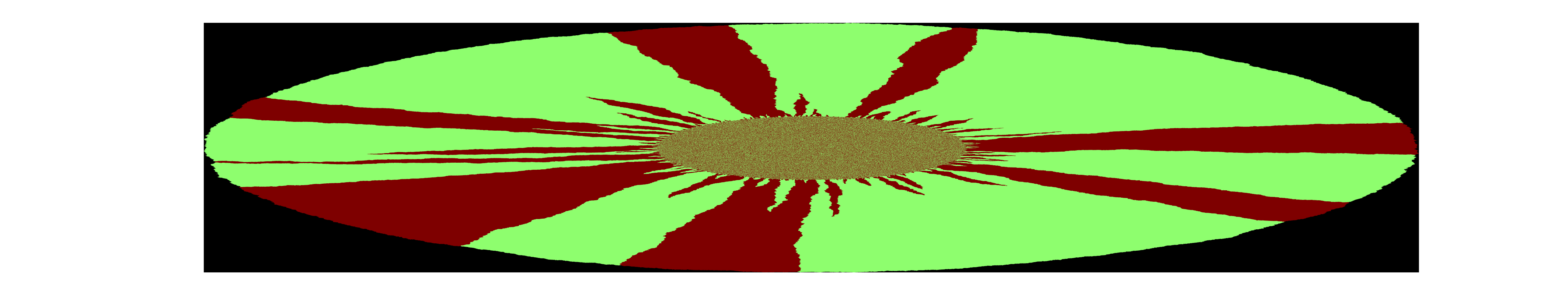}
\end{center}
\caption{\label{fig:1}
Growth patterns of green and red cells grown from a well mixed circular initial population of about $10^6$ cells. Left: Fluorescent image of real experimental data \cite{1} for flourescently labeled E.\ coli bacteria, taken after 84 hours of growth. Right: Data of a single Monte Carlo sample of the lattice based growth model (see Section 2) of comparable inital population size. Both share the same characteristic coarsening patterns.
}
\end{figure}

The motivation for this paper are results in \cite{1}, where the connection between range expansions and random genetic drift is analysed in a simplified experimental setting for populations of microbial systems. A well mixed population of two fluorescently labeled E.\ coli strains is placed at the center of an agar plate containing rich growth medium. Apart from the flourescent label the bacteria are genetically identical. Fig.~\ref{fig:1} (left) shows a fluorescent image of the bacterial population at the end of a growth period of four days. Compactness within the initial habitat and the use of immotile E.\ coli leads to the population evolving only at the leading edge. As the colony grows, cells that loose contact to the colonization front can no longer participate in the colonization process. So the reproducing subset is a small fraction of the total population, chance effects are enhanced and provide the major force for gene segregation via genetic drift. From the initially well mixed population, sectors of single gene alleles emerge, which can either expand or loose contact to the population front. This drives a coarsening process, which leads to a gradual decrease in the number of sectors as a consequence of genetic drift acting at the leading edge of the range expansion. The sectors surviving this annihilation process grow in size at the cost of those left behind, and appear to surf on the wave front \cite{5,2new}. Since the experiment uses immotile E.\ coli on hard agar, there is no noticable change in the population behind the expanding front. Therefore Fig.~\ref{fig:1} provides a frozen record of the coarsening process during colonization. The domain coarsening process can be understood through the fluctuating path of the boundaries separating neighbouring sectors. A domain is frozen out when the leading ends of its boundaries meet and annihilate. A quantitative model and analysis of sector boundaries is presented in \cite{2,korolev09}, where the assumption was made that the fluctuations of the motion taken by the leading tip of the domain boundary are diffusive. However, the authors also discuss that this idealization is not quite correct, since already in \cite{1} the mean squared displacement of the sector boundaries was estimated from experimental data to grow superdiffusively. This superdiffusive behaviour can be attributed to the roughness of the population front, following works on competition interfaces \cite{14} for rough surfaces in the KPZ universality class \cite{13}.

In this paper, we incorporate the superdiffusivity of the sector boundaries and adapt the model introduced in \cite{2}. Based on ideas in that paper, we present a new approach to separate the domain coarsening due to genetic drift and the domain growth due to range expansion. Under an appropriate mapping the circular growth experiment can be understood in terms of annihilating one-dimensional diffusions in a generic linear geometry. This process has been well studied mathematically (see e.g. \cite{masseretal01,munasingheetal06} and references therein) and serves as a prototype for the coarsening dynamics resulting from genetic drift. The geometric influence due to range expansions is completely summarized by the mapping. We argue that this approach is applicable in various situations, and provides a general understanding of the competing effects of genetic drift and range expansions on population fronts. To corroborate our results with good numerical precision, we divise a lattice based spatial stochastic process similar to \cite{korolev09} reproducing the basic features of the experiments in \cite{1}, and use Monte Carlo simulations for sampling. We would like to stress that the purpose of this model and the whole paper is not an accurate quantitative reproduction of experimental data from \cite{1}, but a contribution to the fundamental understanding of the effects of genetic drift on expanding population fronts as a generic emergent phenomenon for a large class of models.

In the next section we provide a description of the model and simulation methods, followed by a detailed numerical analysis of the sector boundaries in Section 3, which leads to an effective mathematical description of their fluctuations. This is used in Section 4 to understand the domain coarsening under range expansions via a mapping to a generic linear geometry, which is our main result. In Section 5 we summarize and discuss the applicability of our approach in more general situations.

\section{Model and simulation method}

A continuous time Markov chain model on a two-dimensional square lattice is used to replicate the basic features of the experiments performed in \cite{1}. Species labeled A and B are used to represent the two distinctly coloured E.\ coli strains. As the strains are otherwise genetically identical, the species A and B reproduce with the same rates $r_{A} =r_{B}$, which we take equal to one. As seen in the experiment \cite{1}, the growth region has a finite width and we approximate this simply by allowing only those cells to reproduce which occupy a position at the colony front. Any descendant of a cell inherits the type (A or B) from its parent and is placed with equal probability on one of the free neighbouring sites. When a cell loses contact to the boundary, it is dropped out of the reproducing subset and becomes part of the frozen interior of the colony. As the expansion increases these cells provide a traceable record of the random chance events that have taken place, which enables us to classify the ancestors of the current reproducing subpopulation.\\
This process is simulated using a Monte Carlo algorithm similar to a model in \cite{korolev09} with random sequential update, which corresponds to the simulation of the jump chain of the Markov model (see e.g. \cite{norris}, Section 2.2). In each time step, one cell is chosen uniformly at random from the reproducing set $E(t)$ with $N(t)=|E(t)|$ members. Then a cell of the same type is put on one of the neighbouring empty sites which is also chosen uniformly at random. The time counter $t\mapsto t+Exp(N(t))$ is updated by adding an exponential random variable with mean $1/N(t)$ which is the waiting time of the Markov chain. As a last step the reproducing set $E(t)$ is updated according to the growth event that just happened. Depending on the geometry the new cell can become a member of $E(t)$, and others may have to be removed if they lost their only empty neighbouring site. For large sizes of $E(t)$ the waiting time is typically approximated by its mean value $1/N(t)$, which gives a very good approximation of the continuous time Markov chain model.\\
We model the growth experiment \cite{1} on the agar plate by initializing an approximately circular subset of the square lattice of radius $r_0$ with A and B cells independently with probabilities $1/2$, leading to a uniform mixture. The initial reproducing set $E(0)$ is then given by the boundary of that circular set, containing of the order of $2\pi r_0$ cells. In Figure \ref{fig:1} we compare a single sample of the Monte Carlo simulation with the experimental picture from \cite{1} where $\pi r_0^2 \approx 10^6$ (i.e. $r_0 \approx 564$), and both show very good qualitative agreement. This suggests that our Markovian model provides a good reproduction of the basic features of the experiment on a large scale, more quantitative evidence will be given later. This is quite remarkable, since we have ignored all microscopic details of the reproduction mode of E.\ coli which is clearly not spatially homogeneous or Markovian (see \cite{1} and references therein). This is a strong indication that the observed segregation is an emergent phenomenon which is to a large extent independent of microscopic details of reproduction, as is discussed in more detail in Section 5.

In our model the size of the population front is basically given by the circumference and grows linearly with the radius. In Section 4 we will show that all statistical properties of this circular expanding process can be fully understood by mapping it onto a linear geometry where the population front has a fixed width. To simulate this we take a strip $\{ 1,\ldots ,w\}\times\Z_+$ of finite width $w\in\Z_+$ with periodic boundary conditions in $x$-direction. At height $h=0$ we initialize the lattice with A and B cells independently with equal probabilities and then apply our simulation algorithm with the restriction of growing only on sites with positive height $h>0$. The initially flat surface will grow in height with local fluctuations building up (see Figure 2), and we call this the linear model/simulation in the following. The sector boundaries resemble a set of annihilating random walks, which leads to a coarsening of the sectors of A and B species analogous to the circular experiment (see Figure 2, right). But due to the finite width $w$ eventually all the boundaries will annihilate and only one sector remains (see Figure 2 left), which is called fixation. The linear geometry has also been realized in a real experiment \cite{1}, by touching the Petri dish with a sterile razor blade that was previously wetted by a liquid mixed culture of flourescently labeled cells.

\begin{figure}
\begin{center}
\includegraphics[height=2in,width=2in]{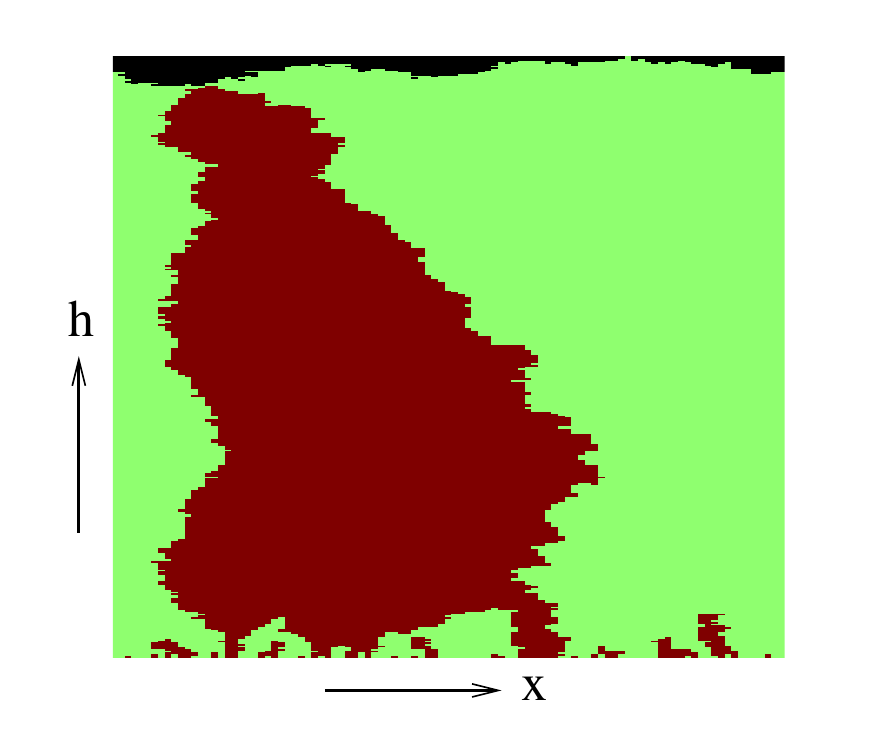}\qquad\includegraphics[height=2in,width=2in]{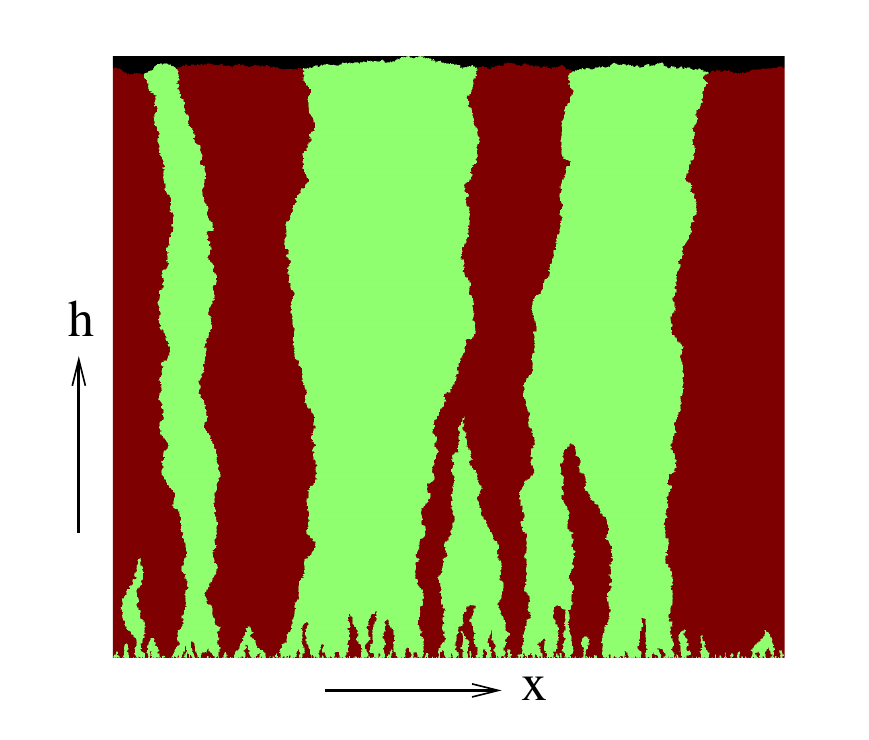}
\end{center}
\caption{\label{fig:3} Sample pictures of the simulation modeling the linear colony experiment, where sites at height $h=0$ are initialised with an A or B cell with equal probability. Fluctuating sector boundaries annihilate when they meet, driving the domain coarsening of allele sectors. The height of the growing frontier is seen to fluctuate as well. Left: For $w=100$ the coarsening of sectors leads to fixation at height $h\approx350$, with only one remaining sector (green in the example shown). Right: For $w=1000$ the width of the strip is too large to reach fixation before the simulation was stopped at height $h\approx1200$. 
}
\end{figure}

\section{First results on a single boundary}

To understand the coarsening behaviour of the genetic domains, we will first study the fluctuations of a single sector boundary and its connection to the roughness of the growing frontier in the linear geometry. We remark that the first results in this section are well known within statistical mechanics. We include a short exposition for completeness and refer the reader to \cite{10,11,12,14} for details.

\subsection{Roughness of the frontier}

\begin{figure}
\begin{center}
\includegraphics[width=4in]{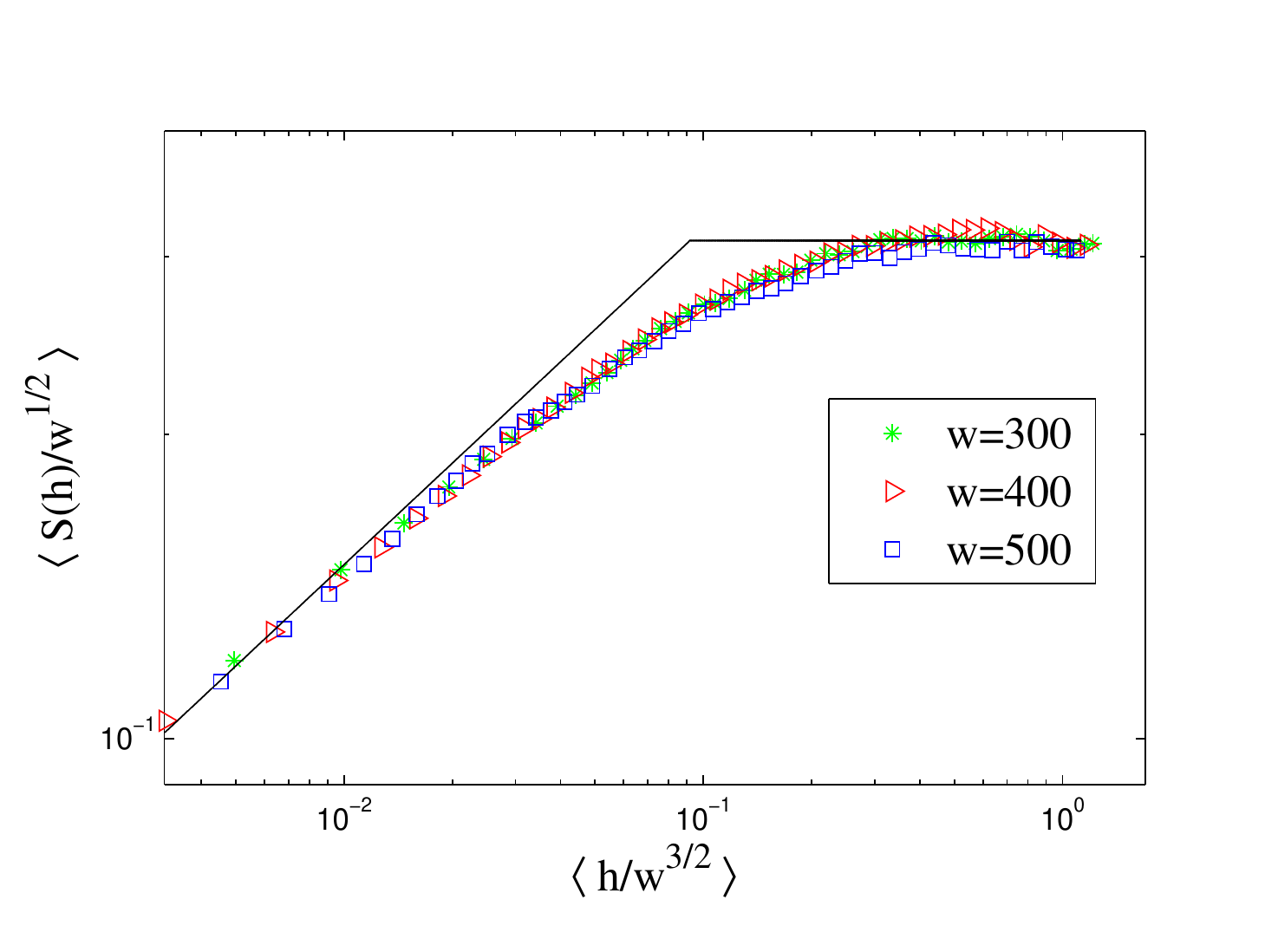}
\end{center}
\caption{\label{fig:4}Double logarighmic plot of surface roughness $S(h)$ as given in (\ref{e1}) for three different width values $w=300$, $w=400$ and $w=500$. $\langle..\rangle$ denotes an average over 500 measurements, data are binned logarithmically to improve the visualization for large values of $h$. The data collapse confirms the scaling prediction (\ref{e3}) of the KPZ universality class for the roughness of the population frontier. The straight lines indicate the initial growth with power $1/3$ and saturation at a constant value.
}
\end{figure}

If we only consider occupied and unoccupied sites and neglect the species, the growth of the population in our lattice based model is well known as the Eden model of type C \cite{10,12}. It is also known, that the surface of Eden clusters belongs to the KPZ universality class \cite{11,13}, which implies fluctuations of the surface height to scale with an exponent $1/3$ as explained below. At any given time, we denote by
$$\bar{h}(t)=\frac{1}{N(t)}\sum_{i\in E(t)} h_{i} (t)$$
the average surface height, where the sum is carried out over heights $h_i (t)$ of the sites in the reproducing boundary $E(t)$, containing $N(t)$ cells. The fluctuations are measured by the roughness, which is defined as the standard deviation of the surface height,
\begin{equation}\label{e1}
S(\bar{h})=\bigg[\frac{1}{N(t)}\sum_{i\in E(t)}(h_{i} (t)-\bar{h} (t))^{2}\bigg]^\frac{1}{2}\ .
\end{equation}
Since $\bar h(t)$ is typically increasing proportional to time $t$, we can see $S(\bar h)$ as a function of the average height, which we will use as a time-like parametrization in the following. Results for the KPZ universality class predict the following behaviour for the function $S(h)$, where write simply $h$ for its argument from now on. When the surface is grown with periodic boundary conditions on a lattice of width $w$ (see e.g. \cite{10}) we have
\begin{equation}\label{e3}
S(h) \sim \left\{\begin{array}{lr} h^{\beta} &\hspace{1cm} \mbox{for $h\ll w^z$} \\w^{\alpha} &\hspace{1cm} \mbox{for $h \gg w^{z}$}\end{array}\right.\ ,
\end{equation}
where the exponents are given by $\alpha =1/2$, $\beta =1/3$ and $z=\alpha /\beta =3/2$ \cite{13}. So for small heights $h$ the roughness $S(h)$ grows like a power law, and saturates for large $h$ to a stationary value depending on the width $w$ of the strip. The crossover occurs at a scale $h\sim w^{3/2}$ and the theory predicts further, that under rescaled coordinates $h/w^{3/2}$ the normalized roughness $S(h)/w^{1/2}$ follows a universal scaling function for all values of $w$. This is confirmed in Figure \ref{fig:4}, where we show data for $S(h)$ averaged over 500 realizations (indicated by $\langle..\rangle$) in a double logarithmic plot. After rescaling we see a data collapse for different values of $w$, and the limiting behaviour (\ref{e3}) for small and large $h$ is indicated by straight lines which fit the data well. This result is of course not new, but confirms the arguments in \cite{1,korolev09} by independent simulations, that the KPZ universality class is indeed the correct framework to study this kind of space limited population growth.

\subsection{Single sector boundaries}

As explained in the Introduction, the superdiffusive behaviour of domain boundaries can be attributed to the roughness of the interface \cite{1,14}. The roughness affects the fluctuations of the sector boundaries, since locally they grow perpendicular to the population front. This leads to an increase in the fluctuations, and to the mean squared displacement of the boundaries growing as $h^{4/3}$ with height \cite{14}. It was noted in \cite{1} that this theoretical value coincides well with preliminary measurements of phase boundary fluctuations from real experimental data. In the following we present simulation data confirming these results, and use them to establish a model for phase boundaries which generalizes the purely diffusive approach in \cite{2,korolev09}.

\begin{figure}
\begin{center}
\includegraphics[width=0.75\textwidth]{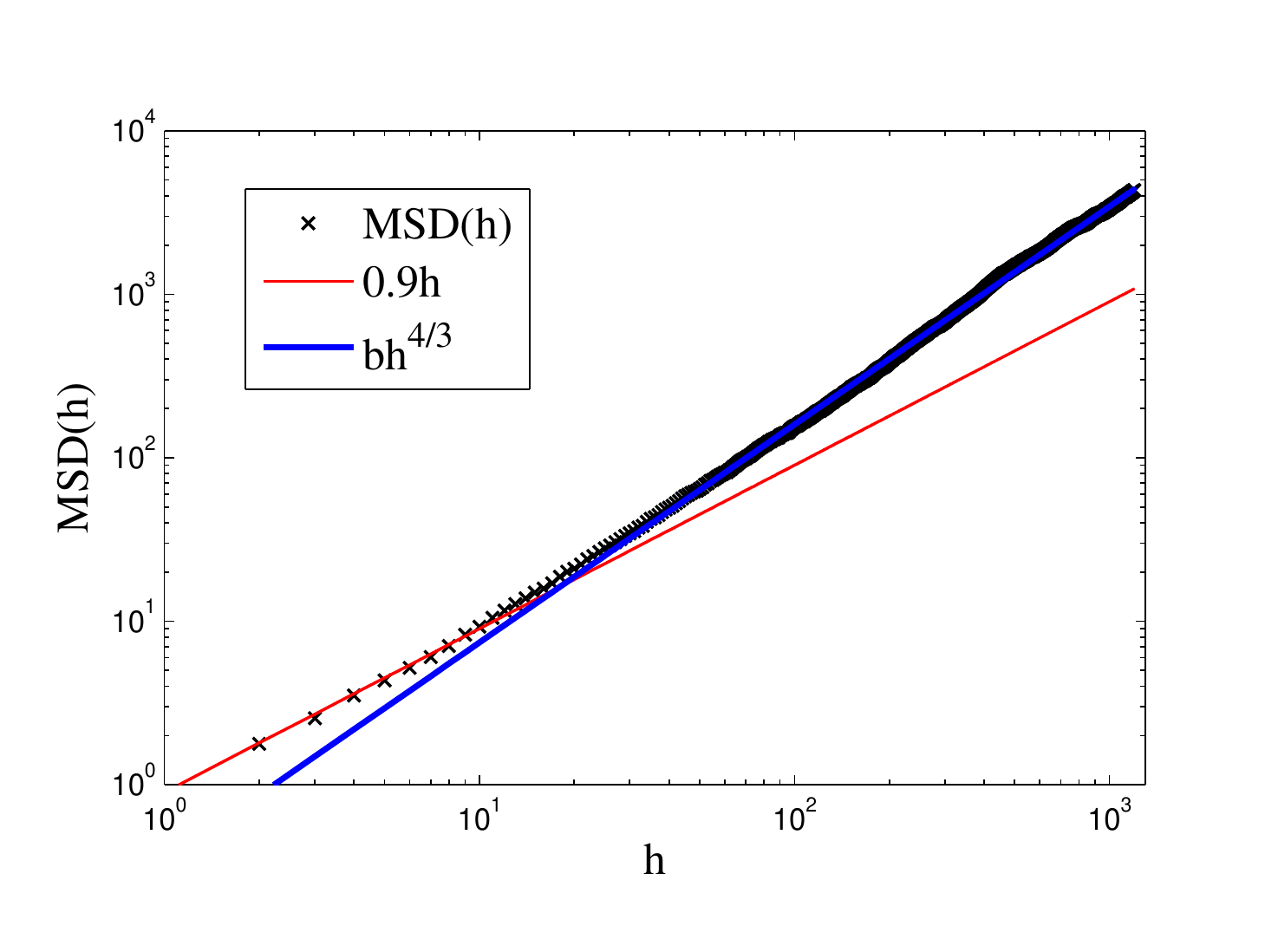}
\end{center}
\caption{\label{fig:7}
Measurements of the mean squared displacement $MSD(h)$ for the sector boundaries (\ref{e5}). The full lines indicate that for small $h$ the behaviour is diffusive $MSD(h)\approx 0.9\, h$ (red) due to the flat initial conditions, while the best fit to the data (blue) of the form $bh^{4/3}$ is $b=0.344$. This describes the data very well in accordance with the theoretical prediction \cite{14}.
}
\end{figure}

We simulate the linear colonial model with the special initial condition of having only species A for $x\in\{ 1,\ldots ,w/2-1\}$ and only species B for $x\in\{ w/2,\ldots ,w\}$, where $w$ is even. Due to periodic boundary conditions, this corresponds to two sector boundaries located initially between sites $w/2-1$, $w/2$ and $w$, $1$. We choose a large width $w=1000$ and stop the simulation at time $t=1000$ well before the fixation time for this system, which is of order $w^{3/2}$ as explained below. Therefore the sector boundaries stay well separated throughout the simulation and can be considered to be independent. 
The path of a sector boundary is composed of single step motions of the tip, which can move up (U), down (D), left (L) and right (R). We analyze 1000 samples of sector boundaries, and our first observation is that the proportion of down steps $\pi_D$ of the tip 
is very small for all the samples, in summary we find
\begin{eqnarray}
	\langle\pi_U \rangle =0.671\ ,\quad\langle\pi_D \rangle =0.005\ ,\quad\langle\pi_L \rangle =0.163\quad\mbox{and}\quad\langle\pi_R \rangle =0.161\ .
\end{eqnarray}
Therefore we can effectively describe the boundaries by one-dimensional lateral displacements $X_h$ as a function of the height $h$, simply by cancelling rare down steps of the boundary tip with subsequent up steps. As expected, the probability of moving left or right is approximately equal, which leads to the boundaries following paths of one dimensional symmetric random walkers with $\langle X_h \rangle =0$. The fluctuations of such paths are determined by the mean square displacement
\begin{equation}\label{e5}
MSD(h)=\langle X_h^2 \rangle\ ,
\end{equation}
which is shown in Fig.~\ref{fig:7} 
in a double logarithmic plot. Following \cite{14}, the KPZ behaviour $S(h)\sim h^{1/3}$ of the surface roughness implies the scaling prediction $MSD(h)\sim h^{4/3}$ for the wandering exponent of the sector boundary, which is also in accordance with the real data in \cite{1}. This is confirmed from our simulations in Fig.~\ref{fig:7} where we show the best fit
\begin{equation}\label{e6}
MSD(h)\approx bh^{4/3}\quad\mbox{with}\quad b=0.344\ ,
\end{equation}
and a line proportional to $h$ for comparison. The second one assuming diffusive fluctuations clearly does not explain the data well except for very small heights, since our simulations start with a flat frontier of zero roughness. The fit with fixed exponent $4/3$ works well for the bulk of the data.  A combined fit of the form $ch^{4/3} +dh$ did not give any improvement over the $bh^{4/3}$ fit, so we did not include it here. The best fit for the exponent gives a value of $\gamma =1.32$ which is consistent with $4/3$ within the accuracy of this measurement. In the following we will therefore use $MSD(h)=bh^{4/3}$ with $b$ as the only fit parameter.

\begin{figure}
\begin{center}
\includegraphics[width=0.49\textwidth]{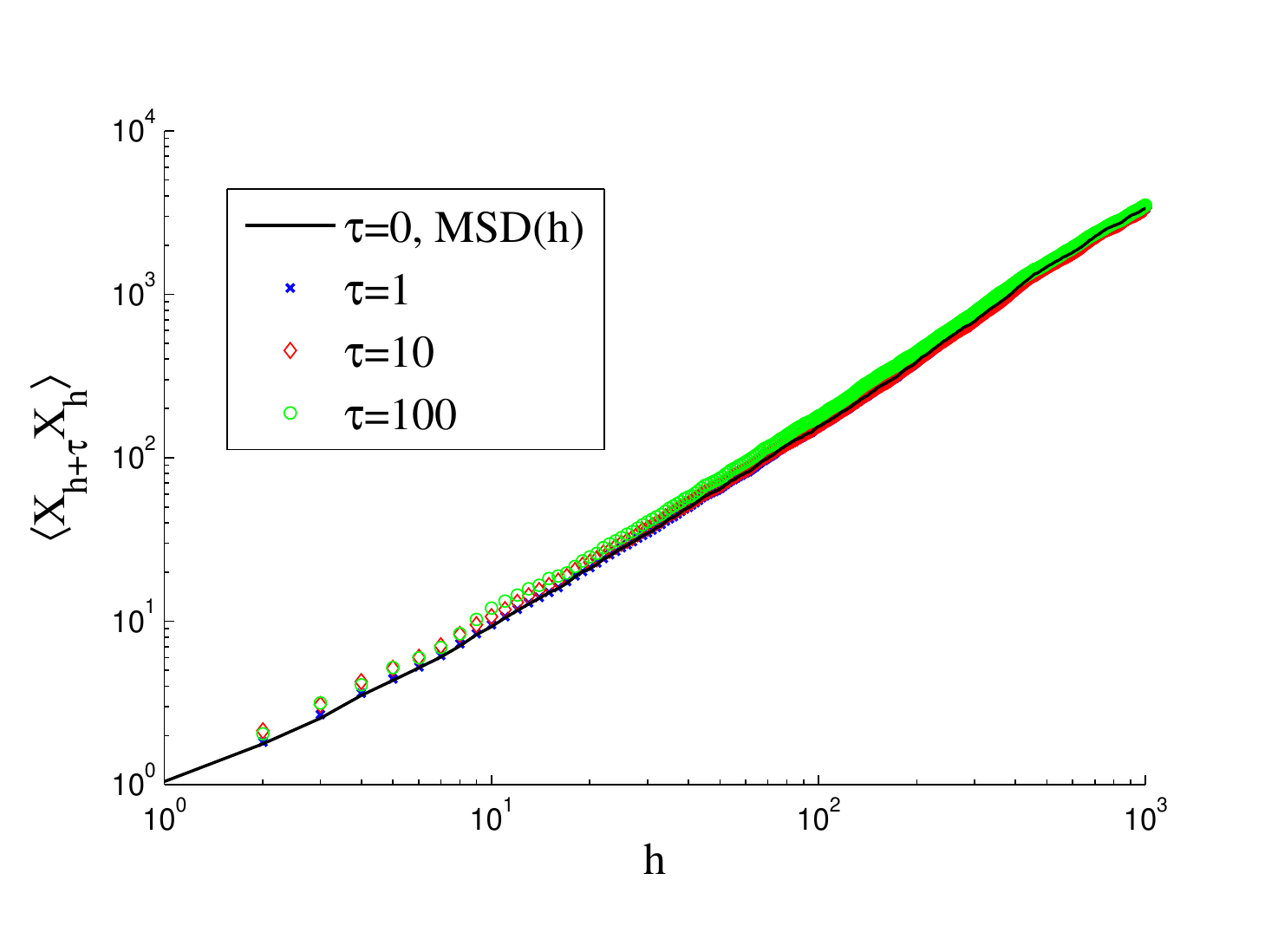}\ \includegraphics[width=0.49\textwidth]{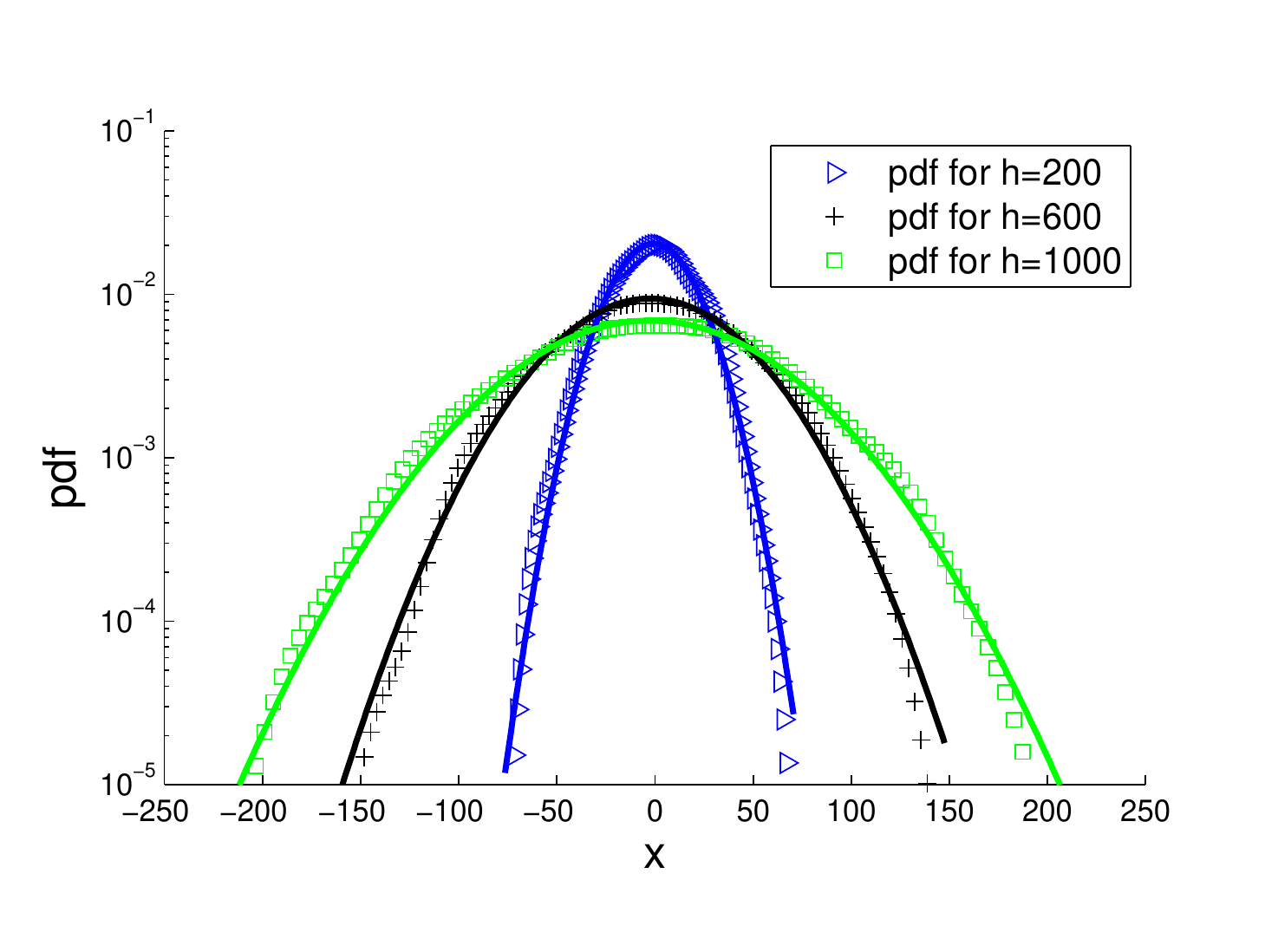}
\end{center}
\caption{\label{fig:6}
Sector boundaries behave like time changed Brownian motions, confirming the model (\ref{bmodel}). Left: We plot the renormalized covariances (\ref{ratio}) as a function of $h$ on a double logarithmic scale for three different values of the lag $\tau$. The data (symbols) are averaged over 1000 realizations and agree well with the expected linear behaviour $h$ (full line). Right: Regularized histograms approximating the pdf of $X_h$ are generated from 1000 realizations, and plotted for three heights $h$ on a logarithmic scale. They agree well with the corresponding Gaussian fit (full lines), deviations for large displacements are due to the finite number of samples.
}
\end{figure}

The superdiffusivity of the boundary is not a result of large jumps as for e.g. Levy processes, but arises from the continuous growth of the surface roughness. Therefore, although superdiffusive, the sector boundaries are still continuous paths on a large scale, which is clearly supported by Fig.~\ref{fig:3}. As a further characterisic, Fig.~\ref{fig:6} (right) shows regularized histograms for the distributions of the displacements $X_h$ for three different values of $h$ which are well approximated by Gaussian probability density functions (pdf's).
The simplest continuous model for the sector boundaries with the correct mean square displacement and Gaussian pdf's is a time-changed Brownian motion. Thus we take
\begin{equation}\label{bmodel}
X_h = B_{MSD(h)}\ ,
\end{equation}
where $(B_t :t\geq 0)$ is a standard Brownian motion with mean square displacement $\langle B_t^2 \rangle =t$ and $B_0 =0$ (see \cite{rogers} Chapter I.1 for details and formal definitions). Under this assumption we expect the covariances
\begin{equation}\label{ratio}
\langle X_h X_{h+\tau } \rangle =\langle B_{MSD(h)} B_{MSD(h+\tau )} \rangle =MSD(h)\ ,
\end{equation}
to be independent of the increment $\tau\geq 0$, which is a standard property of Brownian motions (cf. \cite{rogers} Chapter I.1). 
This is confirmed very well by simulation data shown in Fig.~\ref{fig:6} (left), where the covariances for several values of $\tau$ follow the data for the mean square displacement taken from Fig.~\ref{fig:7}. Therefore our data are in very good agreement with the model (\ref{bmodel}). 

We would like to remark that there are of course many other continuous Gaussian processes such as scaled Brownian motions or fractional Brownian motions as possible models for the domain boundaries. These are characterized by different covariance functions, and none of them showed a better agreement than our model (\ref{bmodel}) in Fig.~\ref{fig:6} (left). The reason we chose (\ref{bmodel}) is that it explains the data very well and is at the same time mathematically easy to treat.

\section{Main results on sector statistics}

\subsection{Simulation data on sector statistics}

In this section we will discuss aspects of the sector distributions, focusing on the average  $\langle M\rangle$ and the second moment $\langle M^2 \rangle$ of the number of sectors $M$ at given times. These statistics are gathered for both the circular and linear colony simulations, where in the first case $M$ is a function of height $h$ and in the second case of radius $r$. Measurements for the circular colony simulations are made for the initial radius values $r_{0}=50$ and $r_{0}=100$, and each simulation is performed until the radius of the population has increased to roughly $7r_0$. For the linear colony simulations we take two corresponding values of the strip width $w=280$ and $564$ so that the initial population fronts have the same size. These are smaller than $2\pi r_{0}$ due to lattice effects.

\begin{figure}
\begin{center}
\includegraphics[width=.75\textwidth]{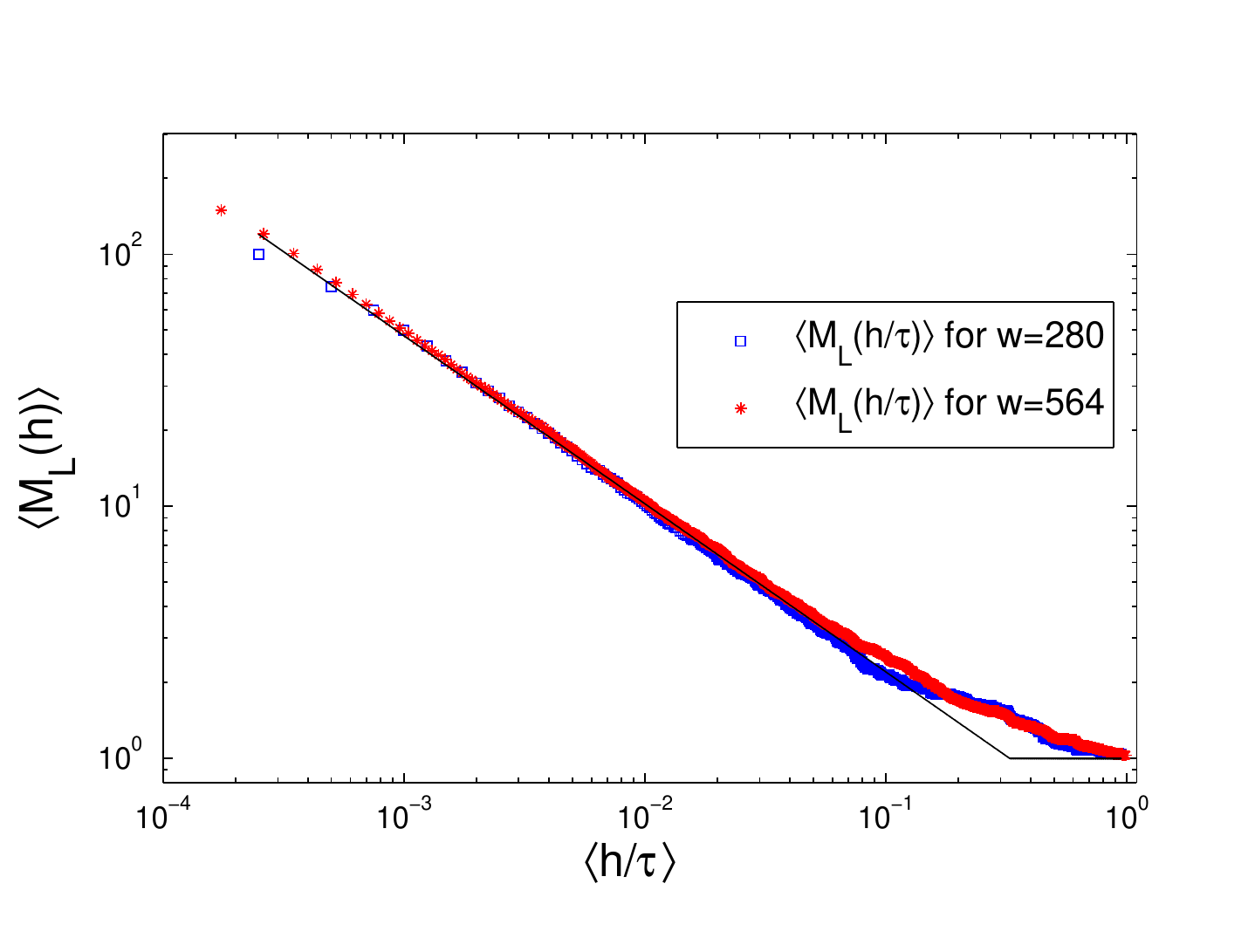}
\end{center}
\caption{\label{fig:8}
Double logarithmic plot of the average number of sectors $\langle M_L\rangle$ in the linear model, as a function of height $h$ rescaled according to (\ref{rescale}). Data points are averaged over 100 samples. The data for the widths $w=280$ and $w=564$ collapse under height rescaling as expected. The straight lines indicate the expected power law $h^{-2/3}$ (\ref{plaw}) and the limiting behaviour due to fixation.
}
\end{figure}

We have seen in the previous section, that the sector boundaries can be modeled by time-changed Brownian motions. It is clear from Fig.~\ref{fig:3} that when two boundaries meet, the sector they enclose will be frozen out and the boundaries annihilate. In the following we will further assume that annihilation is the only interaction between the boundaries, i.e. they move freely until they meet, which is a good first approximation according to our simulation data. For the linear case we are just looking at a set of annihilating Brownian motions in one dimension, which is well understood theoretically \cite{masseretal01,munasingheetal06}. In particular it is known that the average number of sectors decays like $1/\sqrt{ MSD(h)}$ with height $h$. This exact result can be understood heuristically, since the the rate at which boundaries meet and annihilate is proportional to the typical range of a boundary $X_h$ covered up to height $h$, which is given by the root mean square displacement. 
Therefore we expect the number of sectors to decay as
\begin{equation}\label{plaw}
\langle M_L (h)\rangle \sim 1/\sqrt{MSD(h)} \sim h^{-2/3}\ ,
\end{equation}
due to the superdiffusive behaviour, as opposed to $h^{-1/2}$ for standard Brownian motions. This prediction is confirmed by simulation data in Fig.~\ref{fig:8}. We show a double logarithmic plot of the average number of sectors $\langle M_L \rangle$, and the data follow a power law with exponent $-2/3$ very closely in the intermediate scaling regime. For large height $h$ the system leaves the power law regime and saturates, reaching fixation where $\langle M_L (h)\rangle\to 1$ with only one sector remaining. The typical scale $\tau$ to reach fixation is determined by
\begin{equation}\label{rescale}
\sqrt{2MSD(\tau )}=\sqrt{2b}\,\tau^{2/3} =w/2\quad\Rightarrow\quad \tau =\Big(\frac{w}{2\sqrt{2b}}\Big)^{3/2}\ ,
\end{equation}
when two boundaries of maximal initial distance $w/2$ meet. Indeed, rescaling of the $h$-axis by the characteristic scale $\tau$ in Fig.~\ref{fig:8} leads to a very good data collapse for different strip widths $w\approx 280$ and $w\approx 564$, and in both cases $\langle M_L \rangle$ reaches the limiting value $1$ just for $h/\tau =1$.

\begin{figure}
\begin{center}
\includegraphics[width=.75\textwidth]{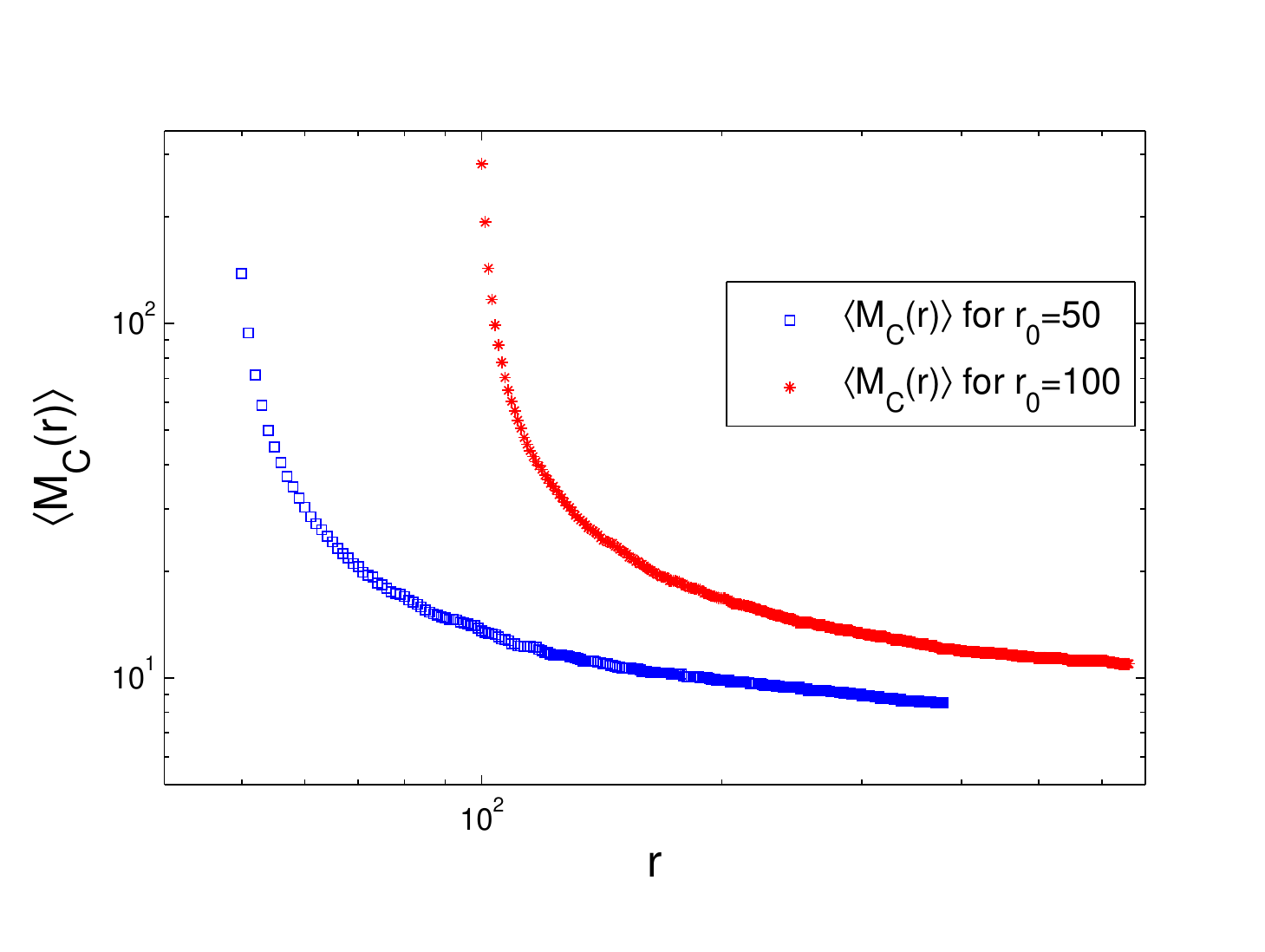}
\end{center}
\caption{\label{fig:9}
Double logarithmic plot of $\langle M_C\rangle$ measured for the circular model as a function of radius $r$ with initial values $r_{0}=50$ and $r_{0}=100$. Data points are averaged over 100 samples. Initially $\langle M_C (r_0 )\rangle =\pi r_0$ and due to the coarsening process $M_{C}(r)$ decays and converges to a stationary value $>1$ which depends on $r_{0}$.
}
\end{figure}

The same data $\langle M_C (r)\rangle$ as a function of radius $r$ are presented in Fig.~\ref{fig:9} for the circular experiment with two different intial radii $r_0 =50$ and $r_0 =100$. We see that the functional form of the decay is not a simple power law in this case, and that for large radius $\langle M_C (r)\rangle$ stays larger than $1$. So in contrast to the linear model which fixates with probability one, the circular system in general does not fixate. Due to the range expansion the size of the colony boundary increases linearly in $r$ and this is faster then the decrease in sector numbers due to coarsening, which is of order $r^{-2/3}$. This slows down the decay in the sector numbers until finally it converges to a stationary value, which depends on the initial radius $r_{0}$. Rather than trying to understand these features of the circular experiment independently (which is of course possible), we will explain in the next section that all its statistical properties are completely determined by the linear model. The geometric effects due to the range expansion can be summarized in a proper mapping between the radius $r$ and heights $h$.

\subsection{Connection between circular and linear model}

To see how the measurements for the two frontier shapes are related, we rescale the statistics of the circular frontier simulation to turn them into a function of an effective height $h(r)$, which compares directly with the statistics of the linear frontier. To do this we first need to find a relationship between a sector boundary $X_{r}$ in the circular colonial simulation with an increasing frontier size, and a domain boundary $X_{h}$ in the linear colonial simulation with a fixed frontier size. Consider an interval $[0, L(r)]$ with periodic boundary conditions, which undergoes a space expansion with increasing $r$, in our case $L(r)=2\pi r$ and initialy we have $L(r_{0})=2\pi r_{0} =w$. In the abscence of diffusion, the relative position of the boundary in the interval should remain fixed, corresponding to a uniform expansion of space. So the rescaled process $Y_{r}=\frac{L(r_{0})}{L(r)}X_{r}$ on the fixed interval $[0,L(r_{0})]$ does not change in time, i.e.
\begin{equation} \label{r1}
dY_{r} = \frac{L(r_{0})}{L(r)}dX_{r} - X_{r}\frac{L'(r)L(r_{0})}{L(r)^{2}}dr=0.
\end{equation}
This will fix the determistic speed of $X_{r}$ due to space expansion and leads to the following SDE for $X_{r}$:
\begin{equation}\label{r2}
dX_{r}= X_{r}\frac{L'(r)}{L(r)}dr +\sigma(r-r_0 )dB_{r} \quad\mbox{for }r\geq r_0\ ,
\end{equation}
where the fluctuations are determined by the model $B_{MSD(r-r_0 )}$ for a single boundary, as established in (\ref{bmodel}). Here we use the incremental form
\begin{equation}\label{r2b}
dB_{MSD(r-r_0 )} =\sigma (r-r_0 )\, dB_r \quad\mbox{with diffusivity}\quad \sigma (r) =\sqrt{MSD'(r)}\ ,
\end{equation}
with a standard Brownian motion $B_{r}$ for $r\geq r_0$ and $B_{r_0} =0$. In the circular case the role of height is taken over by the radius $r-r_0$ relative to the initial condition. Substituting (\ref{r2}) into (\ref{r1}) leads to the following simple SDE for the rescaled process $Y_{r}$,
\begin{equation}\label{r3}
dY_{r}=\frac{L(r_{0})\sigma (r-r_0 )}{L(r)}dB_{r} =\frac{r_0}{r}\,\sigma (r-r_0 )\, dB_r \ .
\end{equation}
Therefore, $Y_{r}$ is a continuous martingale (see e.g. \cite{rogers}, Chapter II.5) and can be written as a stochastic integral
\begin{equation} \label{r4}
Y_{r}=\int_{r_{0}}^{r}\frac{r_0}{s}\,\sigma (s-r_0 )\, dB_{s}\ .
\end{equation}

\begin{figure}
\begin{center}
\includegraphics[width=0.75\textwidth]{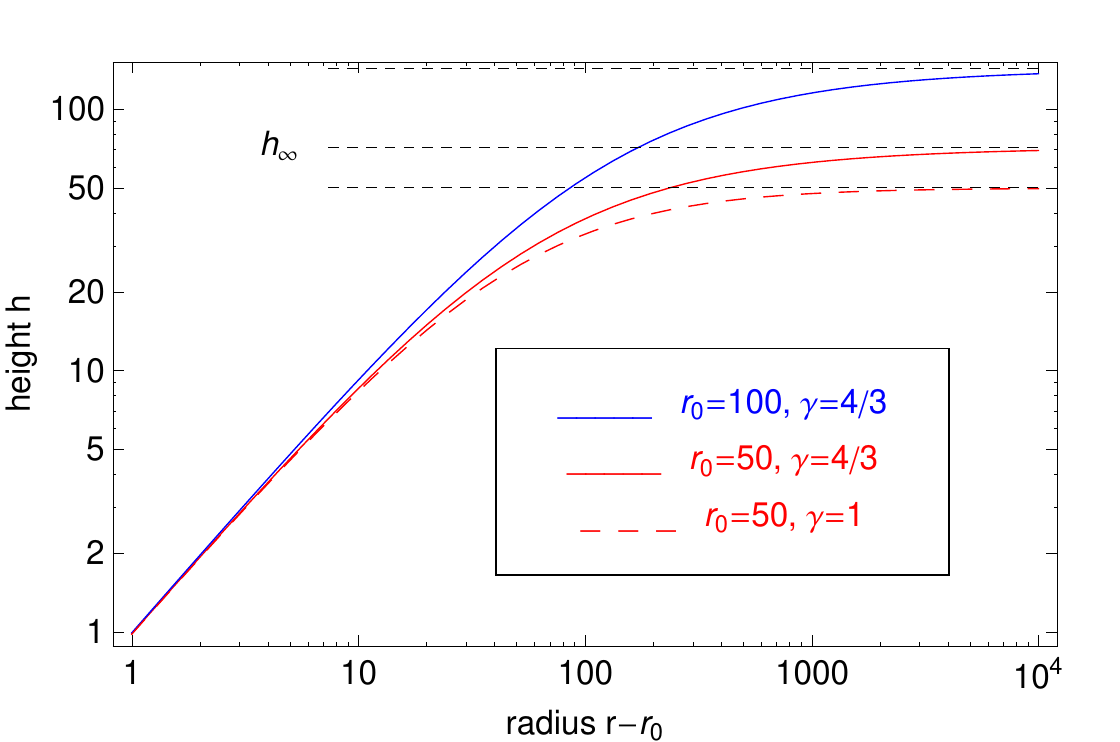}\\
\includegraphics[width=0.4\textwidth]{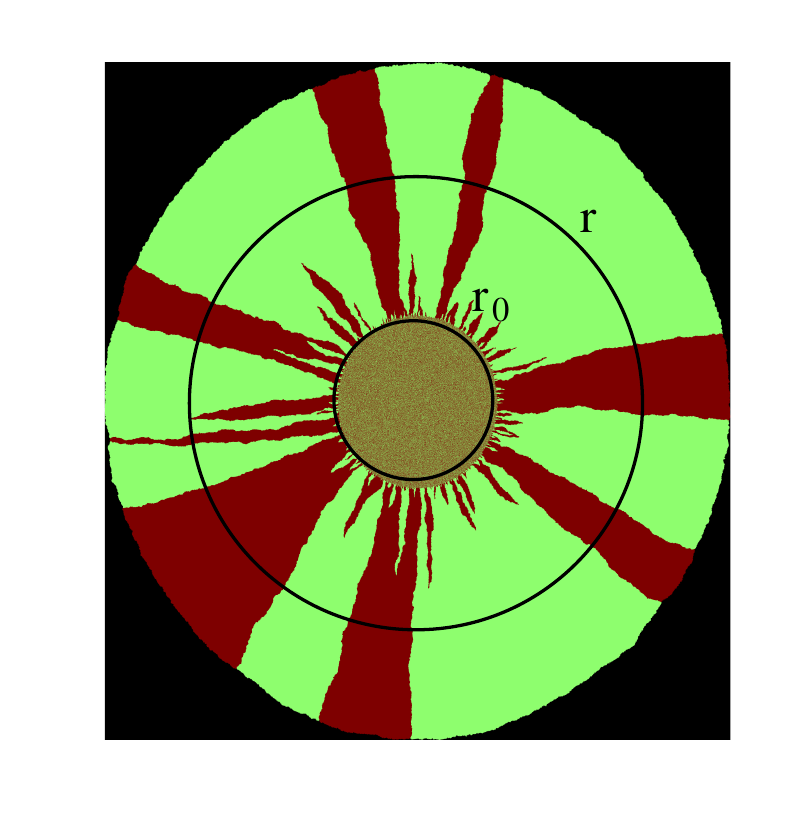}\qquad\includegraphics[width=0.48\textwidth]{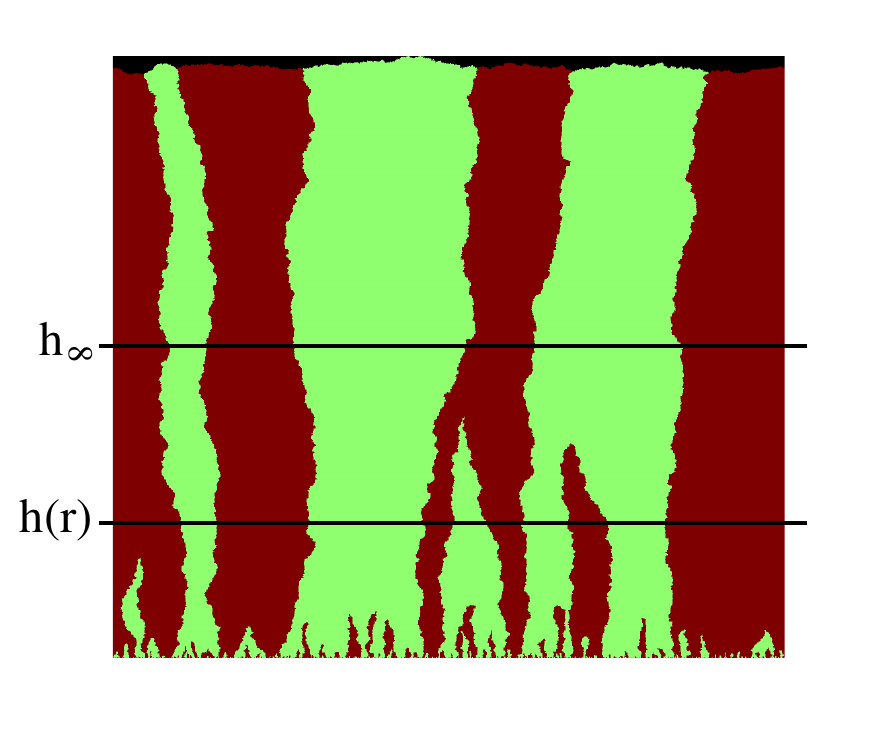}
\end{center}
\caption{\label{fig:10}
The mapping between the circular and the linear model as given by $h(r)$ in (\ref{r7}). Top: Double logarithmic plot of $h(r-r_0 )$ for $r_0 =50,100$ and $\gamma =4/3$ (full lines), and without superdiffusion for $r_0 =50$ and $\gamma =1$ (broken line) for comparison. 
Bottom: Convergence of $h(r)\to h_\infty$ as $r\to\infty$ explains that the circular model (left) exhibits a non-trivial stationary statistic given by the $h_\infty$ line on the right, where typically more than one sector survives.
}
\end{figure}

\begin{figure}
\begin{center}
\includegraphics[width=0.55\textwidth]{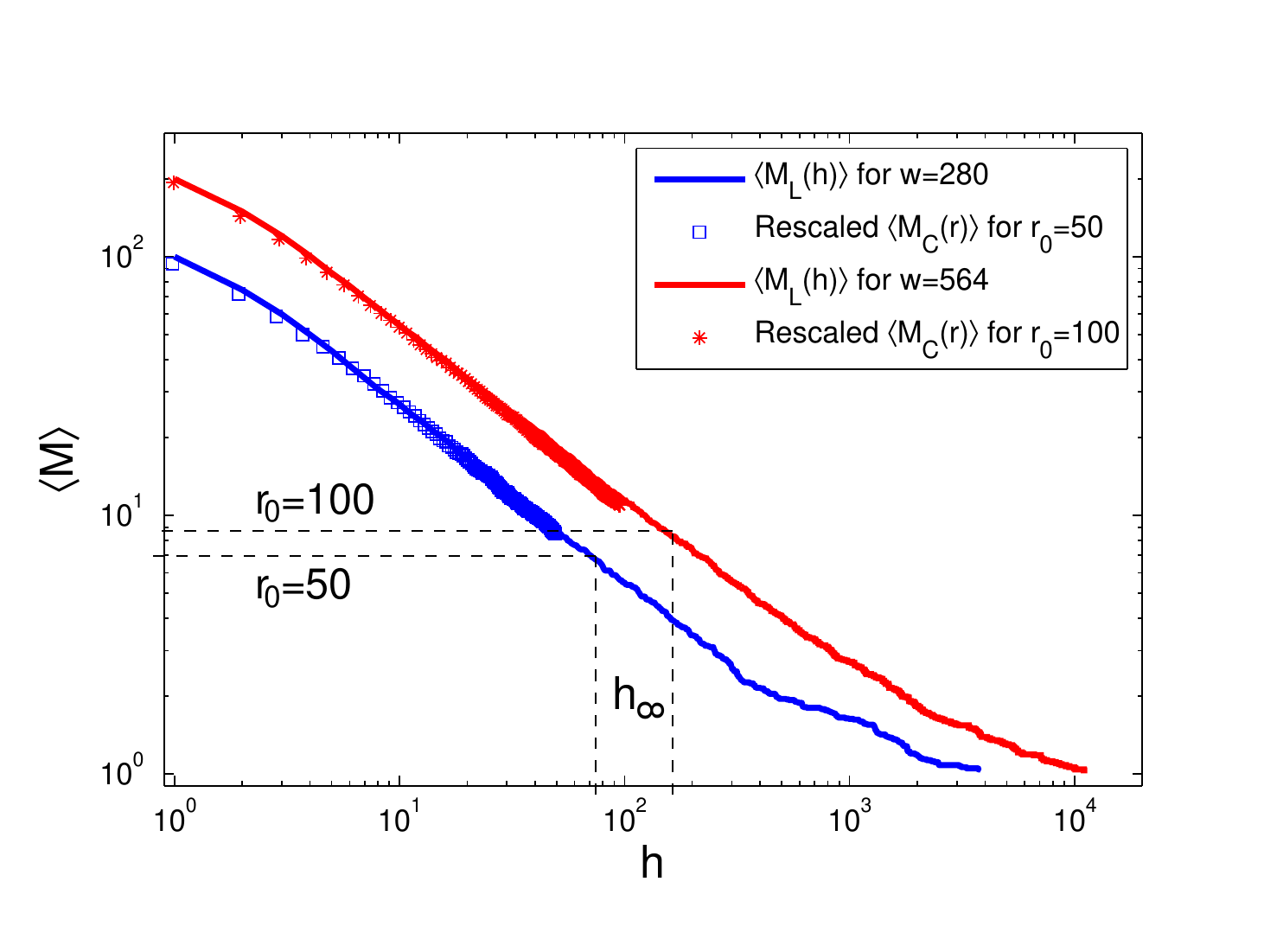}\ \raisebox{5mm}[0mm]{\includegraphics[width=0.44\textwidth]{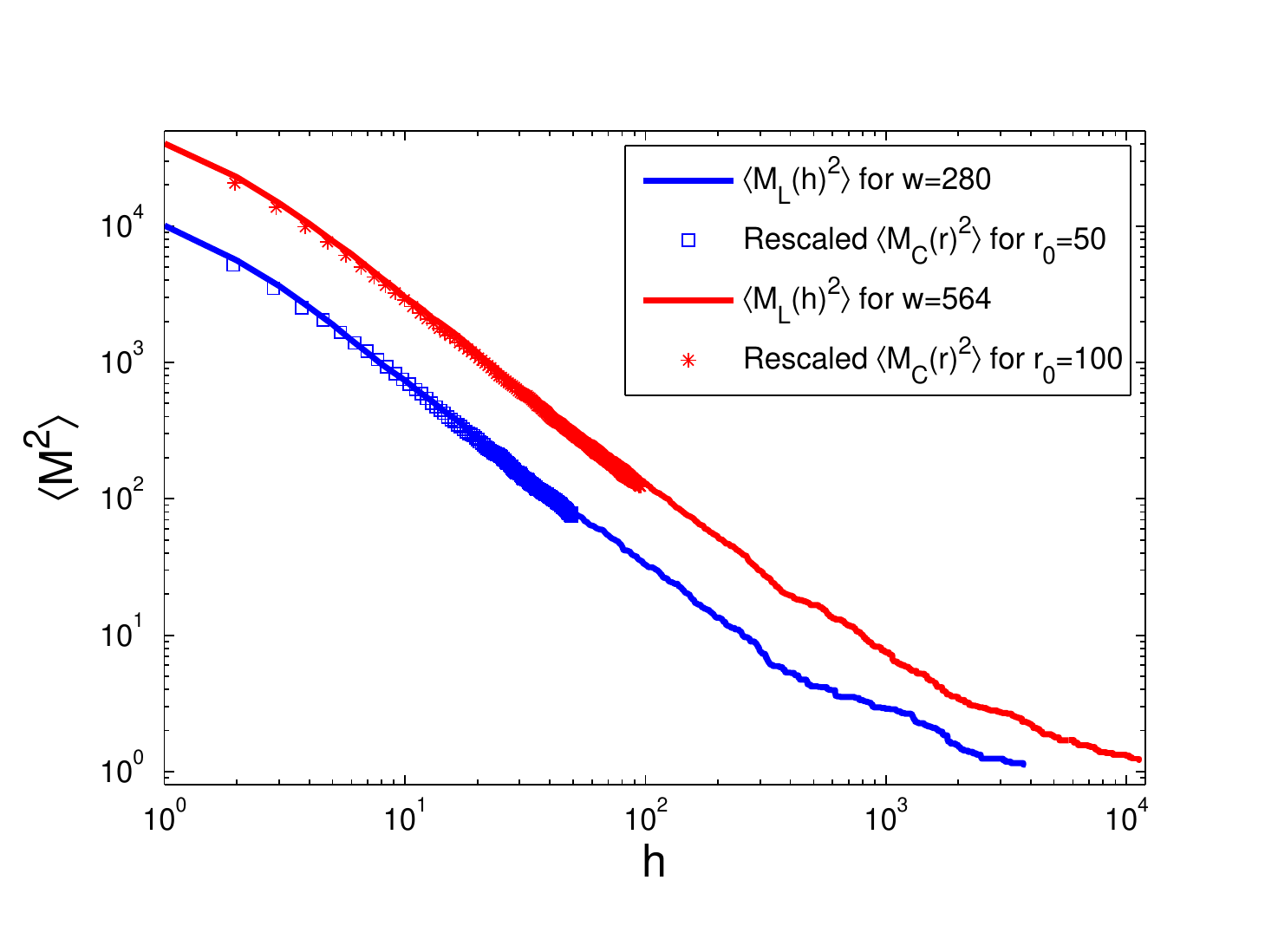}}
\end{center}
\caption{\label{fig:10b}
The mapping between the circular and the linear model applied to our simulation data for $r_0 =50$ and $r_0 =100$ in a double logarithmic plot. Left: The linear data $\langle M_L (h)\rangle$ are shown by full lines, and the corresponding circular data by symbols, where we plot $\langle M_C (r)\rangle$ against $h(r)$ according to (\ref{r7}). The limiting values $h_\infty$ as $r\to\infty$ are indicated by dashed lines and allow to determine the number of sectors $\langle M_C (\infty )\rangle$ for large distances. The data are averaged over 100 realizations. Right: Analogous plot for the second moment $\langle M^2 \rangle$.
}
\end{figure}

\noindent The mean square displacement of the process $Y_r$ is given by the quadratic variation
\begin{equation}\label{r5}
MSD(r)=\langle Y_{r}^{2} \rangle =\int_{r_{0}}^{r}\frac{r_{0}^{2}}{s^{2}}\,\sigma (s-r_0 )^{2}\, ds\ ,
\end{equation}
which is a monotone function of $r$. Also the linear boundary $dX_h =\sigma (h) dB_h$ for $h\geq 0$ is modelled by a continuous martingale as derived in Section 3.2, so both are time changes of standard Brownian motions (see e.g. \cite{rogers}, Chapter II.5). Therefore, under a proper mapping $h(r)$ the circular and linear boundaries are statistically equivalent processes, i.e. $(Y_r :r\geq r_0 )$ and $( X_{h(r)} :r\geq r_0 )$ have the same distribution. The correct time change can be found by equating the mean square displacement (\ref{r5})
\begin{equation}
MSD(r)=bh^\gamma
\end{equation}
with that of a linear boundary as derived in Section 3.2. This leads to the following relationship $h(r)$ between height and radius,
\begin{equation} \label{r7}
h(r) = \bigg[ \frac{1}{b}\int_{r_{0}}^{r}\frac{r_{0}^{2}}{s^{2}}\,\sigma (s-r_0 )^{2} \, ds\bigg]^{1/\gamma} =\bigg[ \gamma\int_{r_{0}}^{r}\frac{r_{0}^{2}}{s^{2}}\, (s-r_0 )^{\gamma -1} \, ds\bigg]^{1/\gamma}\ ,
\end{equation}
where we have used $\sigma (s-r_0 )^2 =MSD'(s-r_0)=b\gamma (s-r_0 )^{\gamma -1}$. This function depends on the initial radius $r_0$ and on the exponent of the mean square fluctuations $\gamma$ which is $4/3$ in our case, but not on the prefactor $b$. For small $r-r_0$ we can neglect the term $(r_0 /s)^2$ in the integral and see that $h(r-r_0 )\approx r-r_0$ is approximately linear. This is to be expected, since close to the initial conditions the linear and circular geometry are locally equivalent and $Y_{r-r_0} \approx X_r$. On the other hand, it is clear that the integral in (\ref{r7}) converges as $r\to\infty$ to a finite number $h_\infty$ depending on $\gamma$ and $r_0$. In fact, $h(r)$ can be written for $\gamma >1$ in terms of the incomplete beta function $B_z (a,b)=\int_0^z s^{a-1} \, (1-s)^{b-1} \, ds$ as
\begin{equation}
h(r)=r_0 \bigg[ \frac{\pi\gamma (1-\gamma )}{\sin (\pi\gamma )}-\gamma B_{r_0 /r}\Big( 2-\gamma ,\gamma \Big)\bigg]^{1/\gamma} \ .
\end{equation}
The first term determines the limit as $r\to\infty$, and for $\gamma =4/3$ its numerical value is given by $h_\infty =1.43\, r_0$. On the other hand, for the diffusive case with $\gamma =1$ the integral (\ref{r7}) can be evaluated easily and we get the even simpler expression
\begin{equation}
h(r)=r_0 \Big( 1-\frac{r_0}{r} \Big)\ \to\ r_0 \quad\mbox{as }r\to\infty\ .
\end{equation}
These results are illustrated in Fig.~\ref{fig:10} (top), where we plot $h(r-r_0 )$ on a double logarithmic scale for the values of interest in our simulations, $r_0 =50,100$ and $\gamma =4/3$. For comparison we also show $\gamma =1$ and $r_0 =50$ and see that neglecting the superdiffusivity of the sector boundaries would lead to a significant change in the mapping.\\
To compare the coarsening process of sector boundaries, note that the circular boundaries $X_r$ meet and annihilate if and only if the rescaled boundaries $Y_r$ do so. Since the latter are statistically equivalent to the linear boundaries $X_h$, the same is true for the statistics of the coarsening process. In fact, using the mapping $h(r)$ the whole circular experiment can be mapped onto and fully understood by the linear one. This is illustrated in Fig.~\ref{fig:10} (bottom), and in Fig.~\ref{fig:10b} (left) we show the rescaled circular data $\big( h(r),\langle M_C (r)\rangle\big)$ and the linear data $\big( h,\langle M_L (h)\rangle\big)$ for the average number of sectors in the same plot. We see a very good agreement for these statistics, and knowledge of the limit $h_\infty =1.43\, r_0$ for $r\to\infty$ can be used to determine the limiting average number of sectors $\langle M_C (\infty )\rangle$ in the circular experiment. This turns out to be around $7$ for $r_0 =50$ and around $8.8$ for $r_0 =100$. Also higher order moments of the number of sectors lead to a very good agreement (see Fig.~\ref{fig:10b} on the right for the second moment), confirming our result of a complete statistical equivalence of the two experiments.

The linear model reaches fixation at heights of order $\tau \sim w^{3/2}$ as derived in (\ref{rescale}) from the finite width $w$ of the strip and the sector boundary fluctuations of order $h^{2/3}$. The circular model undergoes a range expansion where the size of the colony frontier grows linearly with radius $r$ due to the circular geometry. This growth is faster than the fluctuations of the sector boundaries of order $(r-r_0 )^{2/3}$, so that fixation is typically not reached. For the rescaled sector boundaries $Y_r$ the range expansion leads to an effective decrease in the diffusivity $\sigma (r-r_0 )/r$ as seen in (\ref{r3}). This effect is quantitatively characterized by convergence of the function
\begin{equation}
h(r)\to h_\infty \sim r_0 =w/(2\pi )\quad \mbox{as }r\to\infty\ ,
\end{equation}
where the limit is of order $r_0$ with only the prefactor depending on $\gamma$, as we have seen above. The stationary statistics of the circular sectors at large distance is given by statistics of the linear sectors at height $h_\infty \sim w$, so the probability of reaching fixation in the circular model essentially vanishes for large $w$. This can be seen in Fig.~\ref{fig:10b}, where the average number of sectors in the circular experiments are significantly larger than $1$ (which would correspond to fixation).

\section{Discussion}

We have presented numerical evidence, that the sector boundaries of segregation patterns on expanding population fronts can be modelled by time changes of Brownian motion. This lead to a complete understanding of the sector statistics for expanding circular geometries in terms of linear growth models with fixed width, which have been studied in great detail in the mathematics literature. The mapping (\ref{r7}) implies that $h(r)$ converges to a finite limit $h_\infty$ as $r\to\infty$ as long as
\begin{equation}
MSD'(r)/r^2 \sim r^{\gamma -3} \ll 1/r \quad\mbox{for }r\to\infty\ ,
\end{equation}
which holds for all $\gamma <2$. So as long as the motion of sector boundaries is not ballistic (i.e. $\gamma =2$), expansion will dominate the sector coarsening process and there is no fixation for expanding population fronts in a circular geometry. Ballistic sector boundaries can result from different fitnesses modelled e.g.\ by different reproduction rates $r_A$ and $r_B$ for the species. In that case the effects of circular range expansion and coarsening are on the same scale, which leads to interesting competition effects \cite{2,korolev09}. On the other hand, the strength of the range expansion effects can also be affected by non-circular geometries which lead to a non-linear growth of the population front with distance. This could be the result of certain geometrical constraints in the growth medium such as a landscape, which can therefore affect the biodiversity at the population front. We still expect that such systems can be understood by a mapping to the linear experiment analogous to the one presented in this paper. In general, the linear model provides the prototype for the pure coarsening effects due to genetic drift, and the full behaviour is determined by an interplay with geometric effects, which can be summarized in an effective mapping.

The mathematical basis of this argument is that all continuous martingales are time-changed Brownian motions. Other possible models for the sector boundaries such as scaled Brownian motions or fractional Brownian motions are not martingales and the argument as presented here does not strictly apply. Nevertheless, a mapping between different geometries similar to the one presented here should be justifyable at least heuristically also for other boundary models, since the sector boundaries do not exhibit large jumps or interact on macroscopic scales. These conditions are fulfilled in general as long as the species reproduces locally and does not migrate on a time scale faster than that of reproduction. Note also that the sector boundaries inside the population do not have to be static in more general situations. But as long as the range expansion is fast enough, changes due to death, migration or reproduction inside the population will not affect the behaviour at the population front. So our approach is widely applicable for locally reproducing species under range expansions, as long as boundary growth is the fastest relevant process in the system.

According to the well established paradigm of the KPZ universality class \cite{13}, the scaling behaviour of the roughness $S(h)\sim h^{1/3}$ of the population front (\ref{e3}) is largely independent on the microscopic details of the growth rules. This holds as long as spatial and temporal correlation lengths, resulting from these details or external effects from the environment, are small compared to the system size and growth time, respectively. If this is the case, the exponent $\gamma$ of the path of sector boundaries is therefore not affected by details of the mode of reproduction of a single individuum, which will only influence the prefactor $b$ (cf.\ Equ.\ (\ref{e6})). This determines the width of the sectors in the linear geometry in a very simple fashion (cf.\ Equ.\ (\ref{rescale})), whereas our main result, the mapping $h(r)$, is independent of $b$ and therefore applicable to a large class of locally reproducing species. The lattice based simulation model presented in this paper is only a simple caricature of that class, where we have assumed that cells divide in a spatially homogeneous fashion and follow a continuous time Markov process, i.e. their reproduction time has an exponential distribution. Although these assumptions do not hold for E.\ coli, the correlation lengths referred to above are quite small and our simulations are therefore a good coarse grained model on a macroscopic level (cf. Fig.~\ref{fig:1}). It is quantitatively close, but to get the best possible agreement one would have to fit $b$ from a larger set of experimental data. It is of course an intriguing problem to predict the parameter $b$ from the microscopic details of the mode of replication, and we are currently working on a flexible off-lattice simulation model that will allow us to investigate this in detail. A first interesting example for comparison was also reported in \cite{1}, namely that for yeast (S.\ cerevisiae) the sectoring patterns are significantly finer than the ones observed for E.\ coli, corresponding to a reduced mobility of the boundaries and a smaller value of $b$. This could be the result of the larger size of yeast, but temporal aspects of the reproduction are also expected to play a major role.
%

\section*{Acknowledgements}
The authors are grateful to Robin C.\ Ball and Oskar Hallatschek for valuable discussions. This work was supported by EPSRC grant no. EP/E501311/1.


\end{document}